\documentclass[epj]{svjour}
\usepackage{graphicx}
\usepackage{dcolumn} 
\usepackage{bm}
\usepackage{tikz}
\usepackage{epsfig}

\usetikzlibrary{calc,trees,positioning,arrows,chains,shapes.geometric,%
  decorations.pathreplacing,decorations.pathmorphing,shapes,%
  matrix,shapes.symbols}

\tikzset{
  >=stealth',
  punktchain/.style={
    rectangle, 
    rounded corners, 
    draw=black, very thick,
    text width=22em,
    minimum height=1em, 
    text centered, 
    on chain},
  line/.style={draw, thick, <-},
  element/.style={
    tape,
    top color=white,
    bottom color=blue!50!black!60!,
    minimum width=8em,
    draw=blue!40!black!90, very thick,
    text width=10em, 
    minimum height=1em, 
    text centered, 
    on chain},
  every join/.style={->, thick,shorten >=1pt},
  decoration={brace},
  tuborg/.style={decorate},
  tubnode/.style={midway, right=2pt},
}

\begin{document}

\title{Simulation of Flow of Mixtures Through Anisotropic Porous Media
  using a Lattice Boltzmann Model}

\author{M. Mendoza \and F. K. Wittel \and H. J. Herrmann} \institute{
  ETH Zurich, Institute for Building Materials, Computational Physics
  for Engineering Materials, Schafmattstrasse 6, HIF, CH-8093 Zurich,
  Switzerland \email{mmendoza@ethz.ch} }

\abstract{ We propose a description for transient penetration
  simulations of miscible and immiscible fluid mixtures into
  anisotropic porous media, using the lattice Boltzmann (LB)
  method. Our model incorporates hydrodynamic flow, diffusion, surface
  tension, and the possibility for global and local viscosity
  variations to consider various types of hardening fluids. The
  miscible mixture consists of two fluids, one governed by the
  hydrodynamic equations and one by diffusion equations. We validate
  our model on standard problems like Poiseuille flow, the collision
  of a drop with an impermeable, hydrophobic interface and the
  deformation of the fluid due to surface tension forces. To
  demonstrate the applicability to complex geometries, we simulate the
  invasion process of mixtures into wood spruce samples.  } 


\maketitle
\section{Introduction}

Fluid invasion and flow in porous media are ubiquitous phenomena in
nature and technology. In principle, studies on porous media can be
classified by three length scales: the pore scale, the representative
volume element (RVE) scale, and the domain scale \cite{handbook}.
Studies on the pore scale directly model the pore space geometry and
the fluid hydrodynamics \cite{hajoe}. The intermediate sized RVE is
the minimum volume required to characterize the flow through the
medium. Most systems are studied on the RVE scale, using
semi-empirical theories. This is necessary due to complex pore space
geometries. Typical examples are the Darcy, the Brinkman-extended
Darcy, and the Forchheimer - extended Darcy models \cite{handbook}. The
flow in porous media can also be modeled via generalized Navier-Stokes
equations \cite{Tien,Hsu,Nith}, including all necessary terms in one
momentum conservation equation. Unfortunately in the majority of
cases, the analytical solution becomes very difficult and is even
complicated for anisotropic, structured porous media like wood.
Therefore, one depends on numerical methods in this field. The fluid
itself can be quite complicated as well. It can be a miscible or
immiscible mixture of different species, its viscosity can change
locally due to internal chemical reactions like hardening and can
depend on shear velocities in the case of Non-Newtonian
fluids. Examples are adhesive penetration at wood junctures or the
impregnation or sealing of concrete for protective reasons. A
practical model approach for these phenomena is hydrodynamic
dispersion \cite{jacobbear} that treats one fluid, e.g. a solvent, by
the hydrodynamic theory and the others, e.g.  solutes, by diffusion
equations. For complicated anisotropic porous media and fluids,
coupling the diffusion with the generalized Navier-Stokes equations
makes analytical solutions quite difficult.  However, numerical
solutions can be found. The motivation for the present work is
adhesive penetration in wood.

Wood constructions depend on good adhesive bondings, that are the
result of a gluing process, where hardening liquids are pressed into a
porous, anisotropic micro structure. Depending on the penetration and
adhesive hardening characteristics, preferrable interface morphologies
can be obtained \cite{kamke-lee-2007}. The penetration is dominated by
the strong anisotropy of the permeability tensor and the viscosity
evolution of the adhesive. Due to the difficulty of the problem,
penetration simulations with classical continuum mechanics are hardly
feasible.
 
Over the last two decades, lattice Boltzmann (LB) methods developed to
be an alternative to the simulation of partial differential equations
(PDEs). Originally LB methods were developed as discrete realizations
of kinetic models for fluids \cite{n5,n38}. Extensions allow for the
simulation of diffusion \cite{flekkoy}, waves \cite{n101,chop},
magnetohydrodynamics \cite{n39,n6,nmiller}, quantum mechanics
\cite{s93}, and multiphase flows
\cite{ReisPhillips,Mlatva,Daryl,jonathanchin,Xiaowen,andrewk}.
T\"olke et. al. made a first study on binary flows in porous media
\cite{jonasmanfred}. The authors consider two immiscible fluids that
evolve inside a pore scale system by the Navier-Stokes equations with
interaction terms. LB models for incompressible flows through porous
media on the RVE scale were first addressed by Guo et. al.
\cite{preGuo}, who proposed a 2D model including a
nonlinear Forchheimer term, that can be used for isotropic,
heterogeneous porous media.
 
We introduce a 3D LB model that recovers the generalized Navier-Stokes
equations in anisotropic porous media, extended by the diffusion
equation for the solute concentration in fluid mixtures. Consequently
viscosity changes appear, that can be superimposed by local hardening.
Additionally we incorporate the free surface technique
\cite{freesurface} to allow for immiscible fluids.  In this paper,
Sec.~\ref{LBmodel} first describes some LB fundamentals and the
numerical model details for fluid and diffusive particles.
Sec.~\ref{FreeSurfaceSection} shows a brief description of the free
surface technique while Sec.~\ref{algorithm} addresses the
implementation of the algorithm and Sec.~\ref{results} demonstrates
the validity of our model and its application to diverse physical
systems.
\section{3D Lattice-Boltzmann Model for Anisotropic Porous Media}\label{LBmodel} 
We begin with the generalized Navier-Stokes equations for low Reynolds
numbers with mass conservation
\begin{equation}{\label{mass}} 
  \frac{\partial \rho}{\partial t}+\nabla \cdot (\rho \vec{V}) =0 \quad ,
\end{equation} 
and momentum conservation, 
\begin{eqnarray}{\label{momentum}} 
  \frac{\partial \vec{V}}{\partial t}+(\vec{V}\cdot \nabla)\biggl
  (\frac{\vec{V}}{\epsilon} \biggr) =&-\frac{\nabla (\epsilon
    P)}{\rho}+\nu_e\nabla^2\vec{V}+\epsilon \vec{G}\\ \nonumber
  &-\epsilon \nu \tens{K}^{-1}\cdot\vec{V}-\frac{\epsilon}{\rho}\nabla\cdot \tens{S}\quad
  ,
\end{eqnarray} 
with the fluid density $\rho$ and viscosity $\nu$, the volume-averaged
velocity and pressure $\vec{V}$ and $P$. $\epsilon$ denotes the
porosity of the medium, $\tens{K}$ the permeability tensor that can 
be anisotropic, and $\vec{G}$ represents an external force field. The
quantity $\nu_e$ is named the effective viscosity which is not
expected to be the same as the viscosity of the fluid due to the
effects of tortuosity and the dispersion of viscous diffusion flux
\cite{handbook}.  The surface tension tensor $\tens{S}$ can be
expressed by
\begin{equation}\label{eqsurftens} 
  {\bf S}=\frac{2A\tau}{9 |\vec{F}|} \cdot \left( \begin{array}{ccc} - F^2_y-F^2_z & 
    F_xF_y & F_xF_z\\ F_xF_y & -F^2_x-F^2_z & F_yF_z \\ F_zF_x & F_zF_y & 
    -F^2_x-F^2_y 
  \end{array} \right) \quad ,
\end{equation} 
with the color gradient $\vec{F}$$=$$\nabla \rho$, a parameter $A$
that fixes the strength of the surface tension, and the characteristic
relaxation time $\tau$, that is fixed by the effective viscosity
$\nu_e$$=$$\frac{1}{3}(\tau-\frac{1}{2})$ \cite{surfacetension}. Most
of these quantities are related to each other like the pressure
$P$$=$$\rho c^2_s/\epsilon$ depending on the speed of sound of the
fluid $c_s$. In the following we demonstrate how Navier-Stokes
equations are reproduced by LB methods, and how extensions like
external forces, Darcy flow, and surface tension can be added to the
framework.  Diffusion however is treated differently and explained
separately.
\subsection{Model Description for the Generalized Navier-Stokes Equations} 
\begin{figure} 
  \centering \includegraphics[scale=0.5]{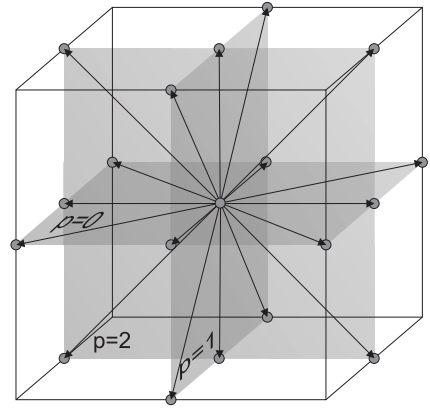}
  \caption{Cubic lattice D3Q19 discretization of the fluid
    velocity. The arrows represent the velocity vectors
    $\vec{v}_{i}^{p}$, where $p$ indicates the plane of location. Note
    that by including the rest vector located in the center of the
    cell, totally 19 directions are defined.}\label{d3q19}
\end{figure} 
The basic principle of LB is that Navier-Stokes equations are not
solved directly on a grid of cells, but via a strongly simplified
particle micro dynamics with discrete Boltzmann equations and
collision rules such as Bhatnagar-Gross-Krook (BGK) \cite{n13}. The two
ingredients particle streaming and collision are calculated along a
certain number of fixed orientations (19 in our case as illustrated in
Fig.\ref{d3q19}) in each cubic cell with lattice constant $\delta
x$$=$$\delta t$. The velocity vectors are denoted by $\vec{v}_i^p$,
where $i$$=$$1...6$, indicates their orientation and $p$$=$$0,1,2$
their reference plane (see Fig.\ref{d3q19}). Each velocity vector
$\vec{v}_{i}^p$ comes along with one assigned distribution function
$f_{i}^{p}$. The density $\rho$ in a cell is the sum over its
distribution functions
\begin{equation}{\label{macros1}} 
  \rho=\sum_{i=1}^6 \sum_{p=0}^2 f_{i}^{p} \quad. 
\end{equation} 
We obtain continuum properties for $\vec{V}$ and $\vec{F}$ via
\begin{equation}{\label{macros}} 
  \rho \vec{V}=\sum_{i=1}^6 \sum_{p=0}^2 f_{i}^{p} \vec{v}_{i}^p \quad ,\qquad
  \vec{F}=\sum_{i=1}^6 \sum_{p=0}^2 \rho(\vec{x}+\vec{v}_{i}^p) \vec{v}_{i}^p
  \quad .
\end{equation} 
 
In order to include the influence of the porous media, of external
forces, and of surface tension on the momentum equation
(Eq.~\ref{momentum}), we need to include additional force terms in the
LB model. Following the work of Guo et.al. \cite{n30}, we add two
force terms $T_{i}^{p}$ and $T_0$ to each of our 19 directions,
including the rest vector in the cell's center:
\begin{eqnarray}{\label{lbe2}} 
  f_{i}^{p}(\vec{x}+\vec{v}_i^p,t+1)-f_{i}^{p}(\vec{x},t)&=\Omega_{i}^{p}(\vec{x},t)+T_{i}^{p}
  ,\\ \nonumber f_{0}(\vec{x},t+1)-f_{0}(\vec{x},t)&= \Omega_{0}(\vec{x},t) +
  T_0\quad .
\end{eqnarray} 
The second ingredient of LB are the BGK collision terms
$\Omega_{i}^{p}$ and $\Omega_{0}$ \cite{n13}
\begin{eqnarray} 
  \Omega_{i}^{p}&=-\frac{1}{\tau}(f_{i}^{p}(\vec{x},t)-f_{i}^{p, \rm  eq}(\vec{x},t))\quad , \\ \nonumber 
  \Omega_{0}&=-\frac{1}{\tau}(f_{0}(\vec{x},t)-f_{0}^{\rm eq }(\vec{x},t))\quad .
\end{eqnarray} 
The subscript $^{\rm eq}$ denotes the equilibrium states for the
distribution functions, which we must find such that the model
reproduces correctly the hydrodynamic equations. $\tau$ denotes again the relaxation time from Eq.~\ref{eqsurftens}.  The force terms $T_{i}^{p}$ and $T_0$
are given by \cite{n30}:
\begin{eqnarray}\label{coeffforce1} 
  T_{i}^{p} &= \biggl ( 1-\frac{1}{2\tau} \biggr)\biggl
  (3(\vec{v}_{i}^p\cdot \vec{F}^{ext}) +
  \frac{9}{\epsilon}(\vec{v}_{i}^p\cdot\vec{V'})(\vec{v}_{i}^p\cdot
  \vec{F}^{ext})\\ \nonumber &-\frac{3}{\epsilon}(\vec{V'}\cdot
  \vec{F}^{ext})\biggr) \quad ,\\ \nonumber &T_0= \biggl (
  1-\frac{1}{2\tau} \biggr) \biggl (-\frac{3}{\epsilon}\vec{V'}\cdot
  \vec{F}^{ext} \biggr) \quad ,
\end{eqnarray} 
with the external force
\begin{equation}\label{forcement} 
  \vec{F}^{ext}=-\epsilon \nu \tens{K}^{-1} \cdot \vec{V'}+\epsilon\vec{G}
\end{equation} 
in our case. Due to the external forces, the velocity $\vec{V}$ needs
to be corrected to obtain $\vec{V'}$ by
\begin{eqnarray}\label{expandV} 
  \vec{V'}=\vec{V}+\frac{1}{2}\vec{F}^{ext} \quad .
\end{eqnarray} 
Unfortunately $\vec{F}^{ext}$ depends on $\vec{V'}$. Then we insert Eq.~\ref{forcement} into Eq.~\ref{expandV} and obtain
\begin{equation}{\label{expandVsub}} 
  \vec{V'}=\tens{\Lambda}^{-1}\cdot \biggl(\vec{V}+\frac{1}{2}\vec{G} \biggr) \quad ,
\end{equation} 
where we define $\tens{\Lambda}$$=$$\tens{I}+\frac{\epsilon \nu}{2}\tens{K}^{-1}$ with the identity matrix $\tens{I}$.
 
We finally write down the equilibrium functions for our system:
\begin{eqnarray}{\label{equilc1}} 
  f_{i}^{p, \rm eq}(\vec{x},t)&=w_i \rho \biggl[1+3(\vec{v}_i^p \cdot 
    \vec{V'})+\frac{9}{2\epsilon}(\vec{v}_i^p \cdot 
    \vec{V'})^2\\ \nonumber &-\frac{3}{2\epsilon}\vec{V'}^2 + \frac{A \epsilon 
      \tau}{\rho} \biggr( \frac{ (\vec{v}_i^p \cdot 
      \vec{F})^2}{|\vec{F}|}-|\vec{F}| \biggl) \biggl ] \quad , 
\end{eqnarray} 
\begin{equation}{\label{equilc2}} 
  f_{0}^{\rm eq}(\vec{x},t)= w_0 \rho \biggl[1-\frac{3}{2\epsilon}\vec{V'}^2
    \biggr] + A \epsilon \tau w_0 |\vec{F}| \quad ,
\end{equation} 
with the known weights $w_{0}=1/3$, $w_{1,2,3,4}=1/36$, and $w_{5,6}=1/18$ for the D3Q19 cells \cite{weightcitation}. The last terms in Eqs.~\ref{equilc1},\ref{equilc2} represent the surface tension. In the continuum limit, the surface tension tensor ${\bf S}$ from Eq.~\ref{momentum} is reproduced.
Note that it can be shown analytically \cite{chapman} that this LB BGK
model recovers the generalized Navier-Stokes equations
Eqs.~\ref{mass},\ref{momentum} in the isothermal and incompressible
limit with surface tension.
\subsection{Lattice Boltzmann model for solute diffusion}
\label{LBDmodel} 
To consider diffusion in an LB framework, we can either define an
additional set of velocity vectors and include their diffusive terms
directly into the equilibrium function (Eq.~\ref{equilc1}), or define
additional equilibrium distribution functions for the diffusion but
work on the same velocity vectors as proposed by Hiorth et.~al.
\cite{difusion} on a D2Q9 cell configuration. In this work, the second
approach was chosen, to simplify a later extension to various species.
We therefore recover the diffusion equation in porous media based on
Ref.~\cite{difusion}, but extend it to heterogeneous porous media.
Starting from the diffusion equation
\begin{equation}\label{eq:diffusion} 
  \frac{\partial C}{\partial t}+\nabla \cdot (C \vec{U}) = \nabla \cdot [D \nabla C] \quad , 
\end{equation} 
with the concentration $C$, the diffusivity $D$, and the velocity of
the mixture $\vec{U}$, Hiorth et.~al.~\cite{difusion} use a D2Q9 cell
configuration and define a distribution function $h_i$ for each
velocity vector. The concentration is calculated via $C$$=$$\sum_{i=1}^9 h_i$ and the equilibrium equations are defined by
\begin{equation}{\label{equildbefore}} 
  h_{i}^{\rm eq}(\vec{x},t)=w_i C \biggl[1+3(\vec{v}_i^p \cdot 
    \vec{U}) \biggr] \quad ,  \quad h_{0}^{\rm eq}(\vec{x},t)= w_0 C \quad , 
\end{equation}
with the weights $w_0$$=$$4/9$, $w_{1,2,3,4}$$=$$1/9$, and $w_{5,6,7,8}$$=$$1/36$ \cite{difusion}. To extend to 3D and to
include the porous medium, we propose modified equilibrium distribution functions on a D3Q19 cell configuration (see Fig.~\ref{d3q19})
\begin{equation}{\label{equild}} 
  h_{i}^{p, \rm eq}(\vec{x},t)=w_i \epsilon C \biggl[1+3(\vec{v}_i^p \cdot  \vec{V^*}) \biggr] \quad ,\quad  h_{0}^{\rm eq}(\vec{x},t)= w_0 \epsilon C \quad , 
\end{equation} 
where $\vec{V^*}$$=$$\vec{V'}+D\frac{\nabla \epsilon}{\epsilon}$. With
the correct weights $w_{0,i}$ for the D3Q19 cell, we evolve
Eq.~\ref{equild} according to the Boltzmann equation with the BGK
collision term $\Omega_{Di}^p(\vec{x},t)$ \cite{n13}:
\begin{eqnarray}{\label{boltzeqD}} 
  &h_i^p(\vec{x}+\vec{v}_i,t+1)-h_i^p(\vec{x},t)=
  \Omega_{Di}^p(\vec{x},t) \quad , \\ \nonumber
   &\Omega_{Di}(\vec{x},t)=-\frac{1}{\tau_D}(h_i^p(\vec{x},t)-h_i^{p,
    \rm eq}(\vec{x},t)) \quad .
\end{eqnarray} 
Now $\tau_D$ defines the relaxation time for the diffusion model with
the diffusivity, $D$$=$$\frac{1}{3}(\tau_D - \frac{1}{2})$. We can prove via a Chapmann-Enskog expansion that the
model reproduces
\begin{equation}\label{diffusionporous} 
  \frac{\partial (\epsilon C)}{\partial t}+\nabla \cdot (\epsilon C \vec{V'}) 
  = \nabla \cdot [\epsilon D \nabla C]
\end{equation}
in the continuum limit.
 
The full model can now reproduce the generalized Navier-Stokes
equations in the continuum limit for anisotropic porous media,
including surface tension and the diffusion-advection of the solute.
To complete the description of the model, we will briefly summarize the free surface technique
\cite{freesurface} used to describe the movement of fluid with a fluid-gas interface.

\section{The Free Surface Technique}\label{FreeSurfaceSection}
If a fluid penetrates an unsaturated porous medium, free surfaces form. To include free surfaces in our LB scheme a free surface technique \cite{freesurface} is necessary. If a cell is only filled by gas with a negligible density, Eqs.~\ref{equilc1} and \ref{equilc2} lead to vanishing equilibrium functions and the simulation becomes unstable. This problem can be solved by classifying fluid cells into three types of cells: the liquid cells, totally filled by liquid fluid, the empty cells entirely filled by gas, and the interface cells that contain both fluids: liquid and gas. The purely liquid filled cell were discussed in the previous section, while gas filled cells can be excluded from the further calculation due to the high density ratio between fluid and gas. However the interface cells need further attention.

Interface cells consist partially of liquid and of gas, determined by the fluid fraction $\lambda(\vec{x},t)$ $= m(\vec{x},t)/\rho(\vec{x},t)$ with the mass $m$ of liquid in the cell located in $\vec{x}$ at time $t$. Using a normalized cell volume, the mass for the fluid cell is equal to the density of the liquid, and for gas cells equal to zero. Analogous to Eq.~\ref{macros1}, we obtain the macroscopic mass $m(\vec{x}, t)$ by summing the mass functions $m_i^p$ over all directions
\begin{equation}\label{eq:massmacro}
  m(\vec{x}, t)= \sum_{i=1}^6 \sum_{p=0}^2 m_i^p \quad .
\end{equation}
The mass functions $m_i^p$ evolve according to
\begin{equation}\label{eq:massevol}
  m_i^p(\vec{x}, t+1)= m_i^p(\vec{x},t)+\sum_{i=1}^6 \sum_{p=0}^2 \Delta m_i^p(\vec{x},t+1) \quad ,
\end{equation}
with
\begin{eqnarray}\label{eq:deltamass}
  \Delta m_i^p(\vec{x},t+1)=
  f_j^p(\vec{x}+\vec{v}_i^p,t)\frac{\lambda(\vec{x}+\vec{v}_i^p,t)+\lambda(\vec{x},t)}{2}
  \\ \nonumber -f_i^p(\vec{x},t)
  \frac{\lambda(\vec{x}+\vec{v}_i^p,t)+\lambda(\vec{x},t)}{2} \quad ,
\end{eqnarray}
where the index $j$ stands for the vector with opposite direction to the vector with index $i$. The interface cells also have associated distribution functions that evolve via the discrete Boltzmann Eq.~\ref{lbe2} identical to those in the fluid cells. These distribution functions are used to calculate the density and the updated fluid fraction $\lambda$.
 
Nevertheless, we still obtain zero distribution functions in the interface cells from Eq.~\ref{lbe2}, due to empty neighboring cells. To avoid this problem, we calculate the evolution of distribution functions of interface cells originating from empty neighboring cells with the modified function
\begin{equation}\label{eq:equilempty}
  f_j^p(\vec{x}, t+1)= f_i^{p, \rm eq}(\vec{x}, t)+f_j^{p, \rm eq}(\vec{x}, t) - f_i^p(\vec{x}, t) \quad .
\end{equation}
During the system evolution, the interface moves, leading to changes in the fluid fraction $\lambda$. If $\lambda$ becomes unity, the cell changes to a liquid cell, while $\lambda \rightarrow 0$ leads to gas or empty cells. To complete the picture all cells neighboring new liquid cells are set to interface cells. To summarize, by classifying cell types and corresponding distribution functions, depending on fluid type, we can capture the evolution of free surfaces. A more detailed description can be found in Ref.~\cite{freesurface}. Finally this technique extends our model to simulations of fluid penetration with gas/liquid interface. Before validation examples are shown, we outline the implementation of the algorithm.
\section{Algorithm Implementation} \label{algorithm} 
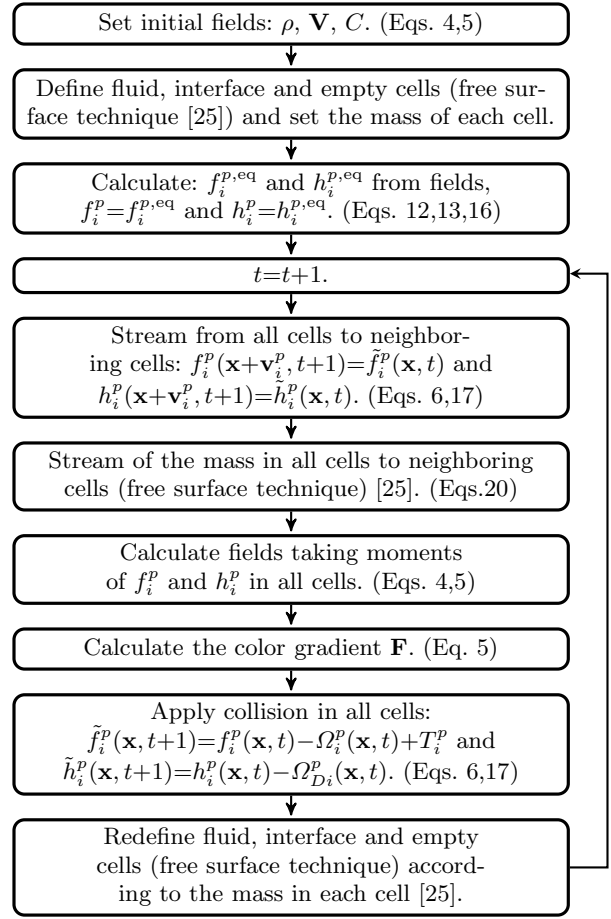
\begin{figure} 
  \begin{center} 
    \small
    \begin{tikzpicture}
      [node distance=.3cm,
      start chain=going below,]
      \node[punktchain, join] (a) {Set initial fields: $\rho$, $\vec{V}$, $C$. (Eqs.~\ref{macros1},\ref{macros})}; 
      \node[punktchain, join] (b) {Define fluid, interface and empty cells (free surface technique  \cite{freesurface}) and set the mass of each cell.}; 
      \node[punktchain, join] (c) {Calculate: $f_{i}^{p, \rm eq}$ and
        $h_{i}^{p, \rm eq}$ from fields, $f_{i}^{p}$$=$$f_{i}^{p, \rm eq}$ and
        $h_{i}^{p}$$=$$h_{i}^{p, \rm eq}$. (Eqs.~\ref{equilc1},\ref{equilc2},\ref{equild})}; 
      \node[punktchain, join] (d) {$t$$=$$t$$+$$1$.};       
      \node[punktchain, join] (f) {Stream from all cells to neighboring cells:
        $f_i^p(\vec{x}$$+$$\vec{v}_i^p,t$$+$$1)$$=$$\tilde{f}_i^p(\vec{x},t)$ and
        $h_i^p(\vec{x}$$+$$\vec{v}_i^p,t$$+$$1)$$=$$\tilde{h}_i^p(\vec{x},t)$. (Eqs.~\ref{lbe2},\ref{boltzeqD})}; 
      \node[punktchain, join] (g) {Stream of the mass in all cells to neighboring cells (free surface technique) \cite{freesurface}. (Eqs.\ref{eq:massevol})}; 
      \node[punktchain, join] (h) {Calculate fields taking moments of $f_i^p$ and $h_i^p$ in all cells. (Eqs.~\ref{macros1},\ref{macros})}; 
      \node[punktchain, join] (e)  {Calculate the color gradient $\vec{F}$. (Eq.~\ref{macros})}; 
      \node[punktchain, join] (i) {Apply collision in all cells:      $\tilde{f}_{i}^{p}(\vec{x},t$$+$$1)$$=$$f_{i}^{p}(\vec{x},t)$$-$$\Omega_{i}^{p}(\vec{x},t)$$+$$T_{i}^{p}$ and
        $\tilde{h}_{i}^{p}(\vec{x},t$$+$$1)$$=$$h_{i}^{p}(\vec{x},t)$$-$$\Omega_{Di}^{p}(\vec{x},t)$. (Eqs.~\ref{lbe2},\ref{boltzeqD})}; 
      \node[punktchain, join] (j) {Redefine fluid, interface and empty cells (free surface technique) according to the mass in each cell \cite{freesurface}. };
      \draw [|-,-|,-stealth, thick,] (j.east)   -- ++(0.5,.0)  -- ++(0,7.9)  -- ++(-0.5,.0);
    \end{tikzpicture}
    \caption{\label{flowchart}  Computational procedure of the LB simulation of flow of mixtures through anisotropic porous media including free surface and tension effects.}
  \end{center}
\end{figure}

To obtain a transparent, portable simulation code, we implemented it object oriented in C++. The general framework of
a simulation is given in Fig.~\ref{flowchart}.
First we set all initial field values like the density $\rho$, the
velocity $\vec{V}$, external force field $\vec{F}^{ext}$
(Eqs.~\ref{macros1},\ref{macros}), and the solvent concentration $C$
inside the mixture. To completely define the system, the porosity
field, the permeability fields, and relaxation times are set. Also the
initial saturation and boundary conditions like periodicity,
pressures, flow rate, or impermeability are imposed. We can define three
types of cells: fluid, interface and empty, depending on the problem.
With these quantities set, we can start calculating the distribution
functions $f_i^p,~h_i^p$ from the equilibrium distribution functions
$f_i^{p, \rm eq}$ and $h_i^{p, \rm eq}$
(Eqs.~\ref{equilc1},\ref{equilc2},\ref{equild}). Now, the time step
can be incremented and system is ready to evolve (see
Fig.~\ref{flowchart}).
 
To follow the evolution of the system, first solve the stream to all
neighboring cells, stream their mass and calculate the macroscopic
variables. Then the color gradient $\vec{F}$ is calculated. Finally, we have to update the type of cells and
increment the time to prepare the system for the next time step. These
steps are looped over all cells in the system. We stop the simulation,
when a stable state is reached.
\section{Model Validation and Calculation of the Flow Profile of
  Adhesives in Wood}
\label{results}
To validate our LB model implementation, we simulate systems with
known analytical solutions that especially address each term that
appears in the generalized Navier-Stokes equations. In particular we
simulate Poiseuille flow through isotropic and anisotropic porous media,
droplet formation, and fluid surface smoothening. Finally, we show an
application for the penetration of hardening mixtures into an
anisotropic, heterogeneous porous medium, demonstrating the capability of
the model to represent adhesive penetration in complex materials like
wood.
\subsection{Single Fluid in Homogeneous, Isotropic Porous Media}\label{homoisoporous}
\begin{figure} 
  \centering \includegraphics[scale=0.35]{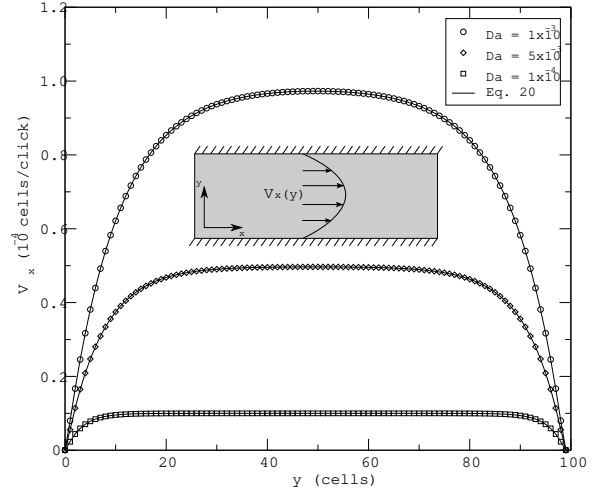} 
  \caption{\label{profile} Velocity profile for a generalized
    Poiseuille flow in a homogeneous, isotropic porous medium for
    various Darcy numbers. The generalized Poiseuille flow is obtained
    when the fluid moves through a channel between two walls driven by a
    constant body force through the porous medium (shaded zone).}
\end{figure}

To show the capability of the model to reproduce the Darcy law in a
simple case, we simulate a generalized Poiseuille flow driven by a
constant force in an isotropic, homogeneous medium between two
infinite plates at distance $L$ (see Fig.\ref{profile}). For this
case, we need neither the surface tension tensor nor the free surface
technique. Therefore just the force terms from the Darcy law, body
force and the viscosity terms on the right hand of Eq.~\ref{momentum}
are active.We impose periodic boundary conditions on all velocity
vectors and null velocity at $y$$=$$0,L$ of our 3D system.  We only
consider movements in the $x$-direction, so the velocity has form
$\vec{V}$$=$$(V_x,0,0)$ with boundary values of
$V_x(x,0)$$=$$V_x(x,L)$$=$$0$.  The analytical solution of this
problem is well known \cite{handbook} by a Brinkman-extended Darcy
equation, and has the form,
\begin{equation} \label{eq:darcy} 
  V_x(y)=\frac{G K}{\nu}\left(
    1-\frac{\cosh[r(y-L/2)]}{\cosh[r L/2]} \right) \quad ,
\end{equation} 
where $r$$=$$\sqrt{\frac{\nu \epsilon}{K\nu_e}}$. 

We set up a system of $3$$\times$$100$$\times$$1$ cells with porosity
$\epsilon=0.3$, relaxation time $\tau=0.6$, density $\rho$$=$$1.0$,
and the body force $\vec{G}$$=$$(0.1,0,0)$.  The permeability $K$ is
fixed by the Darcy number $Da$$=$$K/L^2$ to $10^{-3}$,
$5$$\times$$10^{-3}$, and $10^{-4}$ respectively. We start with a flat
profile and let the system evolve. When the velocity profile reaches a
steady state, the simulation is stopped. Comparing the solution to
Eq.~\ref{eq:darcy} we find excellent agreement with the theory (see
Fig.~\ref{profile}).
\subsection{Anisotropic Permeability} 
In order to validate the model for anisotropic porous media, we study
Poiseuille flow like in Sec.~\ref{homoisoporous}, but with a diagonal
permeability tensor $\tens{K}$, that was fixed to $\tens{K}$ $=$ $10^{-3}
\delta_{xx}+10^{-5} \delta_{yy}$.

\begin{figure}[htb]
 \begin{center}
   \epsfig{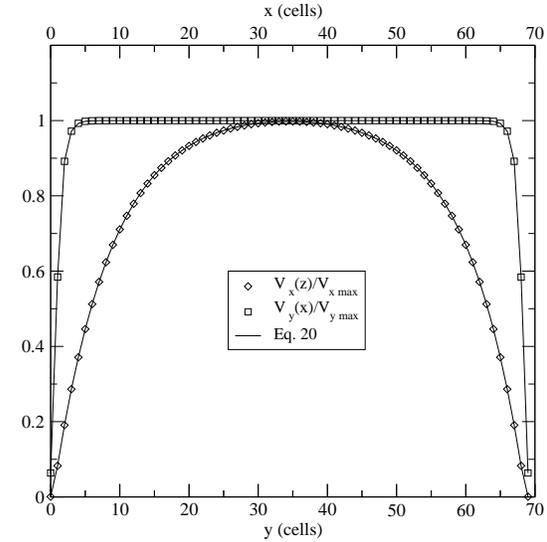}
 \end{center}
 \caption{\label{aniso} Velocity profiles for an anisotropic porous
   medium with diagonal permeability tensor. Squares denote the
   velocity profile of the $x$ component and diamonds represent the
   simulation with flow in $y$-direction ($v_{x,max}=3\cdot 10^{-3}$,
   $v_{y,max}=3\cdot 10^{-5}$).}
\end{figure}
First, we let the fluid move inside the porous medium in $x$-direction
like before with $V_x(y=0,L)$$=$$0$ at the bottom and top. In a second
simulation the fluid moves in $z$-direction with $V_y(x=0,L)$$=$$0$ at
the vertical boundaries. The velocity profiles can be calculated again
by Eq.~\ref{eq:darcy}. In the simulation we used an array of
$70$$\times$$70$$\times$$1$ cells with the same porosity, relaxation
time and density as previously. However the body force is
$\vec{G}$$=$$(0.1,0,0)$ for the first case and $\vec{G}$$=$$(0,0.1,0)$
for the second one.  We run the simulation until the system reaches a
steady state and measure the respective velocity profiles (see
Fig.~\ref{aniso}). The simulation results show that the model
reproduces the Darcy flow in anisotropic porous media accurately.
\subsection{Free Surface Model}\label{chap:freesurf}
As condition for a valid implementation of the free surface technique,
we have to assure, that our model conserves mass. This can be tested
by simulating a freely falling droplet followed by its collision
against an impermeable, hydrophobic wall.

The system consists of a grid of $140$$\times$$140$$\times$$1$ cells.
Initially a droplet with a radius of $8$ cells is placed in the upper
part of the system and an impermeable surface avoids its penetration
through the bottom of the simulation zone at $y$$=$$0$.  The system
parameters are fluid density $\rho=1.0$, relaxation time
$\tau$$=$$0.7$, external gravitational force $\vec{G}$$=$$1\cdot
10^{-6}\cdot(0,1,0)$. The impermeable wall is modelled by setting the
macroscopic velocity in the respective cells to zero. The force terms
from the Darcy law and surface tension are neglected here.

\begin{figure} 
  \centering 
  \includegraphics[scale=0.35]{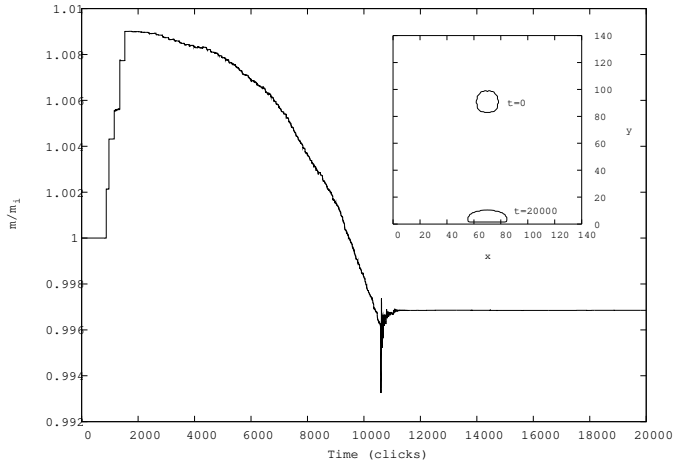} 
  \caption{\label{evolution}Evolution of the total mass of a falling
    droplet that crashes against an impermeable, hydrophobic
    wall. Inset: Snapshots of the droplet at initial time (a) and
    after $20000$ time steps (b).}
\end{figure}

In Fig.~\ref{evolution} we plot the droplet mass as function of
time. We can see that the mass is conserved with a small error of less
than $1$$\%$. Note also that this error is within the range of common
errors that go along with LB simulations. After approximately $10600$
time steps we find an abrupt peak in the total mass due to the
collision with the wall. Note that due to numerical reasons, for a
short time mass is transferred to the boundary cells that represent
the impermeable wall, but is returned in the next time steps, so that
the total mass recovers the value previous to the collision. The
initial and final configurations after $20000$ time steps are shown in
the figure's inset.

\subsection{Introduction of the Surface Tension} \label{chap:surftens}
\begin{figure} 
  \centering 
  \includegraphics[scale=0.38]{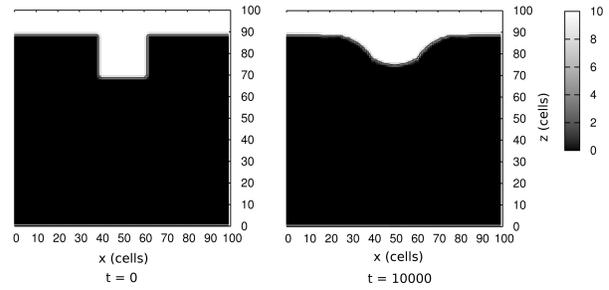} 
  \caption{ Surface smoothing due to surface tension from the initial
    shape (a) to the final shape after $10000$ time steps (b).
    Intensities represent the quantity of fluid in each cell where
    black means empty cells and white completely filled
    ones.}\label{tension}
\end{figure}

Mass conservation alone is not sufficient to describe free surfaces,
rather surface tension needs to be considered. A correct
implementation can be for example tested by simulating the deformation
of the fluid from a rectangular to a circular shape. In this spirit we
implemented a simulation using an array of
$100$$\times$$100$$\times$$1$ cells, with an initially filled, square
fluid zone of size $20$$\times$$20$ cells (see Fig.~\ref{tension}).
The density and relaxation time are identical to the droplet
simulation, and a surface tension parameter $A$$=$$0.001$ was chosen
(see Eq.~\ref{eqsurftens}). The force terms from the Darcy law and
external forces like gravity were switched off for this
simulation. The system evolved until a steady state was reached after
$10000$ time steps. The final shape of the fluid surface smoothened
due to surface tension is seen in Fig.~\ref{tension}.

Summarizing, we showed that the model can reproduce the theory of
single fluids in the case of anisotropic porous media with surface
tension effects and external forces. Note that in our case it is only
possible to apply the free surface technique to liquids with low
Reynold numbers.
\subsection{Penetration of Adhesives in Wood} \label{chap:penetration}
To demonstrate the applicability of our model, we describe the
simulation of the penetration of adhesives into wood which is an
anisotropic porous medium. The adhesives can be treated like two
fluids: a polymer (solute) and a solvent, e.g. water. The polymer
dynamics is governed by the diffusion equation and the solvent by the
generalized Navier-Stokes equations. To consider hardening effects in
the adhesive, we have to consider a local change in time of the
viscosity due to the change of concentration. An additional difficulty
is the description of the anisotropic porous medium wood. We need to
obtain the permeability tensor of wood and to implement a law for
the dependency of the viscosity with the concentration and time.
\paragraph{Wood properties:}\label{chap:woodprops}
In wood, the orthotropic permeability tensor and density is a local
property, depending mainly on year ring geometry, orientation, and
position inside the year ring. Following Refs.~\cite{wittel,Bouriaud}
an empirical expression of the wood density of the type
\begin{equation}\label{wooddensityfunc} 
  \rho(x)=\rho_0 [1+2a x^b \exp(-c x)] \quad ,
\end{equation} 
with the parameters $a$, $b$, $c$, the position $x$, and the minimum
density $\rho_0$ can be made with parameters depending on the type of
wood. For spruce we can take values of $\rho_0$$=$$310$kg/m$^3$,
$a$$=$$\frac{24}{\Delta}$, $b$$=$$0.8$, and
$c$$=$$\frac{9.6}{\Delta}$, where $\Delta$$=$$2.9$mm is the year ring
width \cite{wittel}. To represent several parallel year rings, we use
a periodic function of the form of Eq.~\ref{wooddensityfunc} with
periodic $\Delta$. Since wood consists of cell walls and lumen, we can
estimate the porosity from the density by
\begin{equation}\label{epsilonfunction} 
  \epsilon(x)=1 - \frac{\rho(x)}{\rho_{max}} \quad , 
\end{equation} 
where $\rho_{max}$ is the density of cell walls. Permeability and
porosity are related in the scalar case by \cite{handbook}
\begin{equation}\label{KvsDp} 
  K=\frac{\epsilon^3}{n(1-\epsilon)^2} D_p^2 \quad ,
\end{equation} 
with the characteristic pore size $D_p$ and a constant $n$ that
parametrizes the microscopic geometry of the material taken as
$n$$=$$150$ in our case. In wood, the pore size $D_p$ must also depend
on the wood density, and we propose as a first approximation the
relation
\begin{equation}\label{Dpfunction} 
  D_p(x)=C_2 \left(1 - \frac{\rho(x)}{\rho_{max}} \right)^{\frac{1}{3}} \quad  ,
\end{equation} 
with a constant $C_2$, that can consider the shape of the microscopic
pores \cite{gibson-ashby-99}. Inserting Eqs.~\ref{epsilonfunction} and
\ref{Dpfunction} in Eq.~\ref{KvsDp} we obtain the macroscopic
permeability of a sample as a function of the year ring coordinate
$x$. $C_2$ can now be obtained by integrating Eq.~\ref{KvsDp} over one
year ring and comparing it to literature values for the macroscopic
permeability \cite{Nadia,Patrik} of spruce in radial, longitudinal,
and tangential direction. We finally calculate the required local
permeability tensor in cylindrical coordinates $(r, z, \theta )$, as
\begin{equation}\label{Ktensor}  
  \tens{K}= \left(\frac{\rho_{max}}{\rho(x)}
  \right)^2\epsilon(x)^{\frac{11}{3}}\left( \begin{array}{ccc} 8.2 & 0
    & 0\\ 0 & 51470 & 0 \\ 0 & 0 & 0.82
  \end{array} \right)\times 10^{-10}mm^2 \quad . 
\end{equation} 
\paragraph{Hardening Process and Viscosity Model:} \label{hardening}
To determine the viscosity evolution during the hardening process, we have to
take into account two main factors: the chemical reactions that take
place between the adhesive molecules and the solute concentration. The
first factor causes a viscosity increment with time, while due to the
second one, viscosity changes with local solvent concentrations. In
order to describe these mechanisms, we choose a viscosity model given
by
\begin{equation}\label{nuall} 
  \nu(C,t)=\nu_g [1+a \exp(\alpha t)] \exp(\beta [1-C]) \quad , 
\end{equation} 
where $\nu_g$, $a$, $\alpha$, and $\beta$ are parameters that can be
obtained from experimental hardening curves of adhesives with diverse
solvent concentration.
\paragraph{Simulation of Adhesive Penetration into Wood:}
Embedding the permeability tensor field and the hardening function,
our model is ready to simulate the penetration of adhesives into the
porous wood mesostructure. We use an array of
$90$$\times$$90$$\times$$1$ cells and locate the wood sample in the
zone $y$$<$$60$. The year rings are inclined by $45^\circ$ to the
horizontal plane. We take the fluid density $\rho=1.0$, the parameters
for the porosity, according to Eq.~\ref{wooddensityfunc} with
$a$$=$$0.62$, $b$$=$$0.2$, and $c$$=$$-0.2$, corresponding to spruce
wood \cite{wittel}, and a year ring distance $\Delta$ of $30$
cells. The viscosity model uses the parameters $\nu_g$$=$$0.15$,
$a$$=$$2.05\times10^{-9}$, and $\beta$$=$$10^{-3}$, obtained from
viscosimetric measurements with urea formaldehyde adhesive
\cite{snfreport}. Additionally, we applied a constant external
pressure of $P_{ext}$$=$$0.0441$ that is equivalent to $0.9$mN/mm$^2$
in IS units.
 
\begin{figure} 
  \centering \includegraphics[scale=0.6]{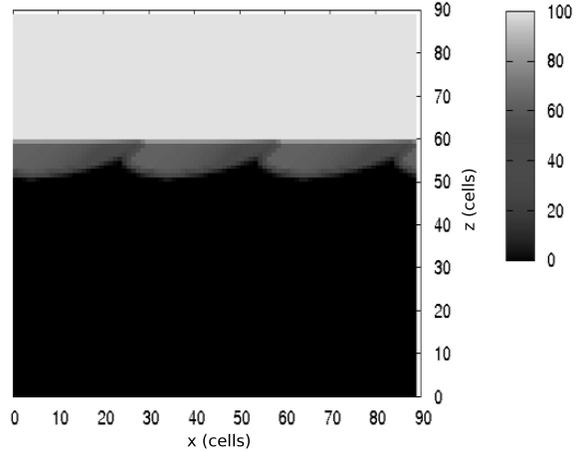} 
  \caption{Fluid profile during the penetration into the wood
    structure. Intensities are identical to Fig.~\ref{tension}. The
    wood sample is located in the region $y$$<$$60$ with a year ring
    inclination of $45^\circ$.}\label{penetration}
\end{figure}

The simulation was stopped when the adhesive could not penetrate
further due to hardening. Fig.~\ref{penetration} shows the penetration
of the fluid into the wood sample. We can see that the model can
simulate complex materials like wood. In some zones the adhesive
penetrates with more speed due to the higher local porosity and the
direction of the movement is correlated with the principal axis of the
permeability tensor.
\section*{Discussion and Conclusions} \label{conclusion} We
introduced a new lattice Boltzmann model to simulate the dynamics
of the flow of mixtures in anisotropic, heterogeneous porous media.
Our 3D model can be applied to many problems in material science and
engineering. It includes a free surface technique to simulate the
invaded fluid profile inside the material structure and a surface
tension term to control the inerface shape dynamics and the forces of cohesion
inside the fluid.
 
The accuracy of the model was tested for a set of simple cases, like
generalized Poiseuille flow in isotropic and anisotropic porous media,
droplet formation and surface smoothing. We found excellent agreement
with the Darcy law for Poiseuille flow. The surface free technique
validated in the simulation of a freely falling droplet and its
collision against an impermeable and hydrophobic surface showed
negligible errors. The surface tension effects were tested by
simulating the smoothing of the fluid profile.  To demonstrate the
applicability of the model to real cases, we implemented a simulation
of the penetration of adhesives into a sample of spruce wood. We
showed that the model can reproduce the dynamics of hardening fluids
in complex materials even when the media are anisotropic and
heterogeneous.
 
The actual model has all the advantages known for the Lattice
Boltzmann method like the minimal use computational resources, fast
algorithms that allow for real time simulations, and the ability for
efficient parallelization. Additionally extensions of the model are
rather straightforward. We hope that the proposed simulation approach
will be useful in order to model complex geometries in heterogeneous
and anisotropic porous media even for complicated fluids like mixtures
of two components.
\begin{acknowledgement} 
The authors are grateful for the financial support of the Swiss
National Science Foundation (SNF) under grant no. 116052.
\end{acknowledgement} 
 
\bibliographystyle{epj}
\bibliography{paper1_submit}

\end{document}